\documentclass{jfm}
\usepackage{natbib}
\usepackage[english]{babel}
\usepackage{graphicx}
\usepackage{amsmath}

\newcommand{\sst}{\scriptscriptstyle}
\newcommand{\Reyn}{\text{\textit{Re}}}
\newcommand{\Stro}{\text{\textit{St}}}

\newcommand{\Frou}{\text{\textit{Fr}}}
\newcommand{\ie}{i.e.\ }

\newcommand{\vect}[1]{\boldsymbol{#1}}

\newcommand{\ui}{\mathrm{i}}
\newcommand{\ud}{\mathrm{d}}
\providecommand{\upi}{\upright{\pi}}

\newcommand{\ms}{\kern.10em\relax}

\newcommand{\pfi}[2]{\ensuremath{{\partial #1}/{\partial #2}}}
\newcommand{\vvb}{\bar{\vect{v}}}
\newcommand{\vxb}{\bar{v}_x}
\newcommand{\vrb}{\bar{v}_r}
\newcommand{\pb}{\bar{p}}
\newcommand{\Zb}{\bar{Z}}
\newcommand{\Tb}{\bar{T}}
\newcommand{\rhob}{\bar{\rho}}
\newcommand{\mub}{\bar{\mu}}

\newcommand{\vxt}{\skew3\tilde{v}_x}
\newcommand{\vrt}{\skew3\tilde{v}_r}
\newcommand{\pt}{\skew3\tilde{p}}
\newcommand{\Zt}{\skew3\tilde{Z}}
\newcommand{\Tt}{\skew3\tilde{T}}
\newcommand{\rhot}{\skew3\tilde{\rho}}
\newcommand{\mut}{\skew3\tilde{\mu}}
\newcommand{\Zs}{Z_S}

\newcommand{\om}{\omega}
\newcommand{\omr}{\omega_\text{r}}
\newcommand{\omi}{\omega_\text{i}}
\newcommand{\kr}{k_\text{r}}
\newcommand{\ki}{k_\text{i}}
\newcommand{\ko}{k_0}
\newcommand{\omo}{\omega_0}
\newcommand{\omor}{\omega_{0,\text{r}}}
\newcommand{\omoi}{\omega_{0,\text{i}}}
\newcommand{\kor}{k_{0,\text{r}}}

\usepackage{xcolor}
\usepackage{soul}

\shorttitle{Diffusion-flame flickering as a hydrodynamic global mode}
\shortauthor{D. Moreno-Boza and others}

\title{Diffusion-flame flickering as a \\ hydrodynamic global mode}

\author{%
D.~Moreno-Boza\aff{1}%
\corresp{\email{dmorenob@eng.ucsd.edu}}, %
W.~Coenen\aff{1}, %
A.~Sevilla\aff{2}, %
J.~Carpio\aff{3}, %
A.L.~S\'{a}nchez\aff{1} %
\and%
A.~Li\~{n}\'{a}n\aff{4}%
}

\affiliation{%
\aff{1}%
Department of Mechanical and Aerospace Engineering, %
UC San Diego, %
La Jolla, CA 92093--0411, USA%
\aff{2}%
Departamento de Ingenier\'{\i}a T\'{e}rmica y de Fluidos, %
Universidad Carlos III de Madrid, %
28911 Legan\'{e}s, %
Spain%
\aff{3}%
ETSI Industriales, %
Universidad Polit\'{e}cnica de Madrid, %
28006 Madrid, %
Spain%
\aff{4}%
ETSI Aeron\'{a}uticos, %
Universidad Polit\'{e}cnica de Madrid, %
28006 Madrid, %
Spain%
}

\begin{document}

\maketitle


\begin{abstract}
The present study employs a linear global stability analysis to investigate buoyancy-induced flickering of axisymmetric laminar jet diffusion flames as a hydrodynamic global mode. The instability-driving interactions of the buoyancy force with the density differences induced by the chemical heat release are described in the infinitely fast reaction limit for unity Lewis numbers of the reactants. The analysis determines the critical conditions at the onset of the linear global instability as well as the Strouhal number of the associated oscillations in terms of the governing parameters of the problem. Marginal instability boundaries are delineated in the Froude-number/Reynolds-number plane for different fuel-jet dilutions. The results of the global stability analysis are compared with direct numerical simulations of time-dependent axisymmetric jet flames and also with results of a local spatio-temporal stability analysis.
\end{abstract}



\section{Introduction}
\label{sec:introduction}

At sufficiently low Froude numbers, jet diffusion flames undergo a bifurcation to a periodic flow state referred to as \emph{flame flicker} \citep{Chamberlin1948}. The associated frequencies observed in laboratory-scale experiments are in the range of 10 to 20~Hz \citep{Roquemore1988}. The role of buoyancy as the driving mechanism was recognized in the early theoretical analysis of \citet{Buckmaster1986}, who postulated that the flickering was associated with a modified Kelvin-Helmholtz instability of the annular flow induced by buoyancy in the envelope of hot gases surrounding the jet flame. By performing an inviscid, parallel flow stability analysis of a simplified self-similar model problem (the so-called infinite candle) they were able to determine an expression for the flicker frequency, which was predicted to vary with the one fourth power of the streamwise distance. This dependence, although weak, was readily recognized as a weakness of the results \citep{Mahalingam1991}. As pointed out by \citet{Buckmaster1986}, a ``detailed viscous stability analysis of the complete flow field'' could help to examine the validity of the results of their simplified study, although they recognized that the suggested analysis was ``a formidable undertaking'' at the time. As a result of the increase in computer power and of the development of robust numerical techniques that have occurred in the intervening time, such an analysis can be performed nowadays with reasonable computational cost, that being the main purpose of the present work.

While the early theoretical work assumed a convective instability \citep{Buckmaster1986}, later experimental observations by \citet{Lingens1996a} and \citet{Maxworthy1999} suggested that the flame flickering phenomenon was associated instead with a globally excited oscillation forced by a region of absolutely unstable flow near the base of the jet exit \citep[see also][]{Cetegen2000}. These findings were later supported by experiments \citep{Juniper2009}, direct numerical simulations (DNS) \citep{Jiang2000,Juniper2009,Boulanger2010} and by local linear stability analyses assuming nearly parallel flow \citep{Lingens1996b,SI_2014}. The present work is different from these previous attempts, in that it employs a linear global stability analysis to study the problem. The method has been used successfully in recent years to investigate the stability of nonbuoyant jet flows, including constant-density jets \citep{Garnaud2013a,Garnaud2013b}, compressible high-speed jets \citep{Nichols2011}, and light jets at low Mach numbers \citep{Lesshafft2015,Coenen2016}. The global instability of reacting jets has been considered recently by \citet{Qadri2015}, who studied the buoyancy-free lifted flame investigated earlier by \citet{Nichols2008} and \citet{Nichols2009} using a combination of DNS and local linear stability analysis. All of the previous linear global stability analyses of jet flows have considered buoyancy-free conditions. The method is to be employed below to examine buoyancy-induced flickering of axisymmetric laminar jet diffusion flames. The study provides the critical conditions at the onset of the linear global instability as well as the Strouhal number of the associated oscillations in terms of the governing parameters of the problem.

An important aspect of jet-flow instability concerns the applicability of spatio-temporal linear stability analyses for the predictions of the critical conditions at the onset of the global instability. When the flow is sufficiently slender, in that the resulting eigenmodes are much shorter than the jet development region, then the assumption of nearly parallel flow becomes accurate and the critical conditions can be identified from the analysis of the region where the flow is absolutely unstable, as shown by \citet{Lesshafft2007a}. This slenderness condition is satisfied in buoyancy-free jet flows, for which the eigenmodes scale with the jet radius, which is much smaller than the jet development length for the moderately large values of the Reynolds number that characterize the onset of the instability. For instance, local linear stability analyses of light gaseous jets \citep{Coenen2008,Coenen2012} have shown to give predictions in agreement with those of DNS \citep{Lesshafft2007} and of global stability analyses \citep{Lesshafft2015,Coenen2016}. This is in contrast with the buoyancy-induced flickering flames investigated below, for which the eigenmodes will be seen to scale with the flame length, rather than with the jet radius. Under those conditions, the quasi-parallel assumption no longer holds and predictions based on the local linear stability analysis become necessarily inaccurate, with resulting critical Froude numbers at the onset of the instability that are off by a factor exceeding two, as shown below.

As observed clearly in flow visualizations of jet flames with nearly uniform exit velocity profiles \citep{Roquemore1988}, the flickering mode, characterized by large toroidal vortices surrounding the flame, is accompanied by a Kelvin-Helmholtz instability of the shear layer surrounding the fuel jet leading to the formation of an inner train of small discrete vortices. To focus attention on the flickering phenomenon, our analysis will purposely preclude the emergence of these shear instabilities by considering only cases in which the fuel-feed velocity profile is parabolic, an appropriate boundary condition for sufficiently long fuel injectors. Also, unlike previous authors~\citep{SI_2014}, who used in their stability analysis a detailed flow field description including finite-rate chemistry and advanced molecular-transport models, we choose to employ instead a simplified flow model that retains all relevant aspects involved in the hydrodynamic instability leading to flame flicker while neglecting secondary effects that complicate unnecessarily the description, thereby facilitating both development of fundamental physical understanding and extraction of parametric dependences. For instance, since the variations of density and transport properties in combustion flows are mainly associated with the temperature changes induced by the chemical heat release, a constant average molecular weight will be employed when writing the equation of state and the different transport coefficients will be assumed to be independent of the composition, while their temperature dependence will be approximated by a power law. A Fickian description with unity Lewis numbers will be used for the diffusion velocity of the reactants. Furthermore, we shall consider nonpremixed jet-flame configurations in which the rates of the chemical reactions involved in the fuel-oxidation process are sufficiently fast for the burning rate to be diffusion controlled~\citep{LVL_2015}. Under these conditions, the resulting nonpremixed flame remains anchored in the vicinity of the injector rim and the interaction between the envelope of hot gases surrounding the jet flame and the gravitational acceleration leading to the onset of the flickering mode can be investigated by using the limit of infinitely fast reaction, with the composition and temperature described in terms of a single passive scalar, the so-called mixture-fraction variable. Consideration of finite-rate chemistry is necessary in stability analyses of lifted flames, such as that performed recently by \citet{Qadri2015}.

The paper is structured as follows. The nondimensional equations and boundary conditions are presented in \S\,\ref{sec:formulation}, which is followed in \S\,\ref{The global linear stability analysis} by relevant numerical results, including sample spectra and transition diagrams in the controlling-parameter plane. Comparisons of the predictions of the global stability analysis with results of DNS of unsteady axisymmetric flows are presented in~\S\,\ref{sec:comp}. A local spatio-temporal stability analysis of the transverse profiles of the base flow is performed in~\S\,\ref{sec:local}; the results are seen to significantly overpredict the critical Froude number, thereby underscoring the limited predicting capability of local analyses for buoyancy-induced flickering. Finally, concluding remarks will be offered in \S\,\ref{sec:conclusions}.


\section{Problem formulation}
\label{sec:formulation}

\begin{figure}
    \centering
    \includegraphics[width=0.60\textwidth]{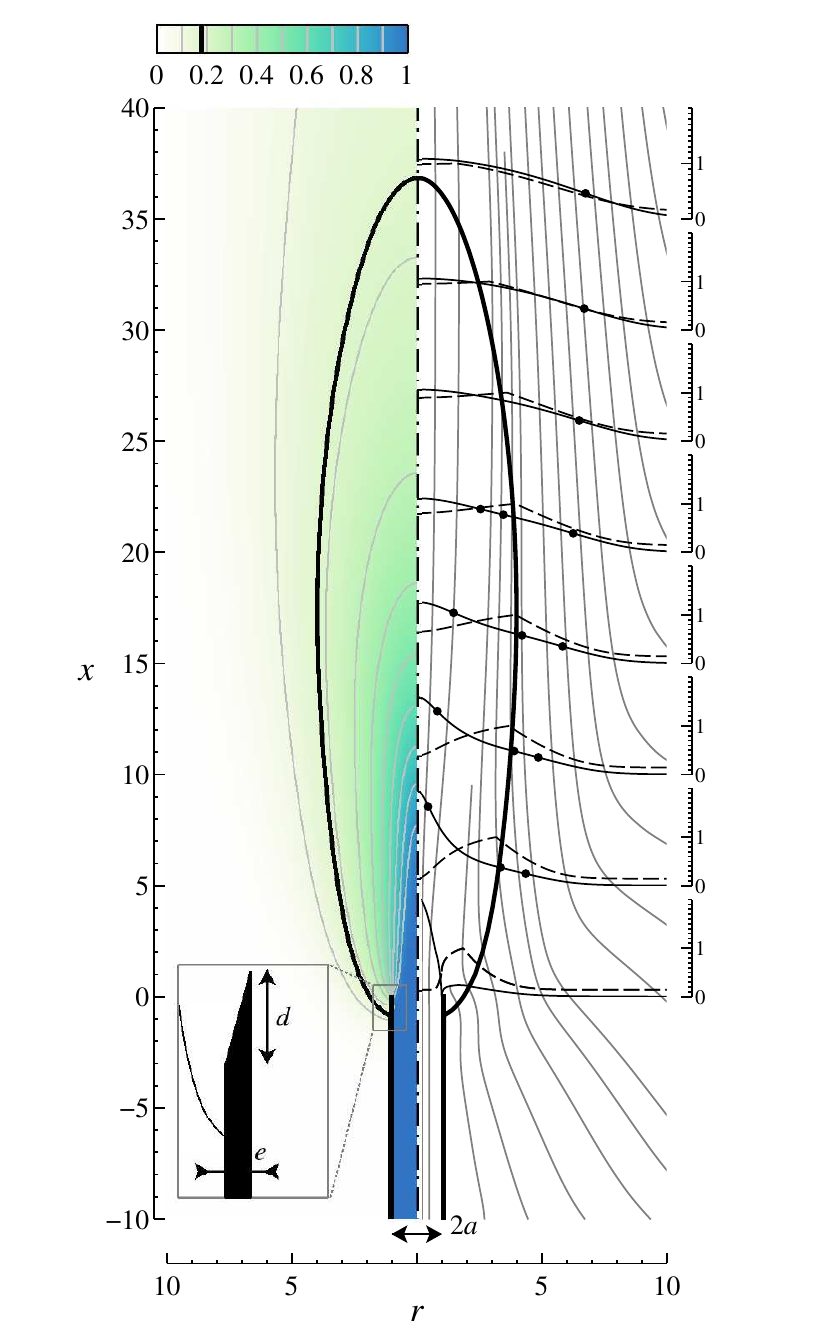}
            \caption{Base-flow isocontours of~$\bar{Z}$ (left-hand side) and streamlines (right-hand side) together with radial profiles of~$\bar{v}_x$ (solid curves) and~$\bar{T}/(\gamma+1)$ (dashed curves) at $x=(0,5,10,15,20,25,30,35)$ for $\Pran=0.7$, $S=4.62$, $\gamma=6$, $\Reyn = 100$ and $\Frou = 300$. The dot on the velocity profiles indicates the location of the inflection points. The thick solid line represents the stoichiometric flame surface $\bar{Z} = Z_S$, where $\bar{T} = 1+\gamma$.}
    \label{fig:sketchbaseflow}
\end{figure}

As indicated in figure~\ref{fig:sketchbaseflow}, the configuration analyzed includes a vertical fuel jet discharging upwards through an injector of inner radius $a$ into an infinite air atmosphere. The specific geometry investigated here involves a thin injector of thickness $e \ll a$. To minimize wake effects, the rim of the injector is knife-like sharpened as indicated in the inset of figure~\ref{fig:sketchbaseflow}. For the numerical integrations shown below the injector wall thickness and the slenderness ratio of the wedge tip were selected to be $e/a=10^{-3}$ and $d/e=20$, respectively. Smaller values of $e/a$ and larger values of $d/e$ were used in sample integrations to ensure that the results were independent of these two geometric parameters, so that the solution given below is representative of infinitesimally thin injectors.

For generality, the analysis considers  dilution of the fuel with an inert gas, with $Y_{\sst{\rm F,0}}$ denoting the fuel mass fraction in its feed stream, while $Y_{\sst{\rm O_2,A}}=0.232$ is the oxygen mass fraction in air. In the description, focused on the fluid mechanical aspects of the flow, we adopt the one-step irreversible overall reaction ${\rm F} + s \, {\rm O_2} \rightarrow (1+s) \, \text{Products} + q$, according to which the unit mass of fuel reacts with a mass $s$ of oxygen, releasing in the process an amount of energy $q$. The above representation of the underlying stoichiometry for the oxidation of the fuel embodies the two fundamental thermochemical parameters involved in nonpremixed combustion~\citep{LVL_2015}, 
\begin{equation}
S =\frac{s Y_{\sst{\rm F},0}}{Y_{\sst{\rm O_2},A}} \quad {\rm and} \quad \gamma = \frac{q Y_{\sst{\rm F},0}}{c_p T'_0 (1+S)}, \label{Sgamma_def}
\end{equation}
the former representing the mass of air that one needs to mix with the unit mass of the gaseous fuel stream to generate a stoichiometric mixture and the latter being the corresponding dimensionless temperature increment resulting from the adiabatic combustion of that mixture. Here, $T'_0$ is the initial temperature of the feed streams, assumed to be equal for the fuel jet and for the surrounding air atmosphere, and $c_p$ represents the specific heat at constant pressure, taken to be constant in the following analysis. Typical values for undiluted hydrocarbon-air flames initially at normal ambient temperature are $S_u = s/Y_{\sst{\rm O_2},A} \simeq 15$ and $\gamma_u \simeq 6-7$. Diluting the fuel stream with an inert gas to give a fuel mass fraction $Y_{\sst{\rm F},0}<1$ in its feed stream has a direct effect on the value of $S = Y_{\sst{\rm F},0} S_u$, but a much more limited effect on the heat-release parameter, as can be seen by writing the second expression in~\eqref{Sgamma_def} for $S_u \gg 1$ in the approximate form $(\gamma_u-\gamma)/\gamma_u \simeq (1+Y_{\sst{\rm F},0} S_u)^{-1}$, which indicates that significant variations of $\gamma$ require extremely dilute fuel mixtures such that $Y_{\sst{\rm F},0} \sim S_u^{-1}$. 

In the limit of infinitely fast reaction adopted here the reaction takes place in an infinitesimally thin layer, outside of which the chemical-equilibrium condition $\hat{Y}_{\sst{\rm F}} \hat{Y}_{\sst{\rm O_2}} = 0$ applies, with $\hat{Y}_{\sst{\rm F}} = Y_{\sst{\rm F}}/Y_{\sst{\rm F},0}$ and $\hat{Y}_{\sst{\rm O}} = Y_{\sst{\rm O_2}}/Y_{\sst{\rm O_2},A}$ representing the fuel and oxygen mass fractions normalized with their values in their respective feed streams. The reaction-rate terms in the conservation equations for energy and species appear as Dirac-delta distributions located at the flame, which becomes in this limit an infinitesimally thin surface attached to the injector separating a near-axis region without oxygen from a fuel-free outer atmosphere~\citep{BS_1928}. For equidiffusive reactants, \citet{Sh_1948} and~\citet{Ze_1949} showed how the computation can be facilitated by the introduction of conserved scalars satisfying transport equations, obtained by combinations of the species and energy conservation equations that eliminate the chemical-source terms. Two conveniently normalized forms of these passive scalars are the mixture fraction and the excess enthalpy, defined as
\begin{equation}
Z = \frac{S \hat{Y}_{\sst{\rm F}}- \hat{Y}_{\sst{\rm O}}+1}{S+1} \quad {\rm and} \quad H=T-1+\gamma (\hat{Y}_{\sst{\rm F}}+ \hat{Y}_{\sst{\rm O}}-1), \label{definitions}
\end{equation}
where the nondimensional temperature $T$ has been scaled with $T'_0$. The mixture fraction is defined to be zero in the air stream and unity in the fuel stream, respectively, whereas at the flame, where both reactants appear in zero concentrations, $Z$ takes the stoichiometric value $Z_S=1/(S+1)$. On the other hand, the excess enthalpy is defined to be zero in both feed streams, so that when the injector walls are adiabatic, the case considered here, the solution for the associated transport equation reduces to $H=0$ everywhere in the flow field, thereby facilitating the description. The piecewise-linear expressions
\begin{subequations}
\label{rels}
\begin{align}
    &\hat{Y}_{\sst{\rm F}} = 0,
        \quad
        \hat{Y}_{\sst{\rm O}} = 1-\frac{Z}{Z_S},
        \quad
        T-1= \gamma \frac{Z}{Z_S};
        & {\rm for} \;
        0\le Z \le Z_S, 
         \label{rel_1}\\
    &\hat{Y}_{\sst{\rm O}} = 0,
        \quad
        \hat{Y}_{\sst{\rm F}} = \frac{Z-Z_S}{1-Z_S},
        \quad
        T-1= \gamma \frac{1-Z}{1-Z_S};
        & {\rm for} \;
        Z_S\le Z \le 1,
        \label{rel_2}
    \end{align}
\end{subequations}
obtained from the definitions~\eqref{definitions} with use made of the equilibrium condition~$\hat{Y}_{\sst{\rm F}} \hat{Y}_{\sst{\rm O_2}} = 0$ and of the result $H=0$, provide the reactant mass fractions and temperature in terms of $Z$. Evaluation of the expressions for $T$ at $Z=Z_S$ indicates that the temperature at the flame surface is everywhere equal to the stoichiometric adiabatic flame temperature $T=1+ \gamma$, a known result of the infinitely fast reaction limit that holds in adiabatic configurations with unity Lewis numbers of the reactants.

The problem reduces to that of integrating the continuity and momentum equations together with the transport equation for $Z$, which are written in the dimensionless form
\begin{align}
    \frac{\p \rho}{\p t} + \nabla \cdot (\rho \vect{v}) &= 0,
    \label{cont} \\
    \rho \frac{\p \vect{v}}{\p t} + \rho \vect{v} \cdot \nabla \vect{v}
        &= -\nabla p + \frac{1}{\Reyn} \nabla \cdot \bar{\bar{\tau}}
        + \frac{1}{\Frou} (1-\rho) \vect{e}_x, \label{mom-eq} \\
    \rho \frac{\p Z}{\p t} + \rho \vect{v} \cdot \nabla Z
        &= \frac{1}{\Reyn \, \Pran} \nabla \cdot
            \left[\rho D_{\sst{\rm T}} \nabla Z \right], \label{Z}
\end{align}
where $\Pran=0.7$ is the Prandtl number and
\begin{equation}
\Reyn=\frac{\rho'_0 U_0 a}{\mu'_0} \quad {\rm and} \quad \Frou=\frac{U_0^2}{g a} \label{ReFr_def}
\end{equation}
are the Reynolds number and the Froude number for the jet flow, respectively, with $\rho'_0$ and $\mu'_0$ representing the density and shear viscosity in the feed streams.
The jet radius $a$ and the average jet velocity $U_0=\dot{m}/(\upi a^2 \rho'_0)$ based on the fuel mass flow rate $\dot{m}$ are used to scale the problem. The development employs cylindrical coordinates $\vect{x}=(x,r)$ centered at the injector exit plane with an associated velocity vector $\vect{v}=(v_x,v_r)$; the streamwise coordinate $x$ pointing against the gravity vector $\vect{g}=-g\vect{e}_x$. 

In the low-Mach number approximation utilized here, the pressure variations can be neglected in the first approximation when writing the equation of state, which therefore reduces to $\rho T=1$
when the additional assumption of constant molecular weight is adopted to achieve maximum simplification, with $\rho=\rho'/\rho'_0$ denoting the dimensionless density. Furthermore, in this low-Mach number limit, the viscous-stress term proportional to the second viscosity coefficient can be incorporated in the definition of the variable $p$ that represents in~\eqref{mom-eq} the pressure difference from the unperturbed ambient distribution. Correspondingly, the resulting viscous-stress tensor reduces to $\bar{\bar{\tau}}= \mu \left(\nabla \vect{v} + \nabla \vect{v}^{\sst{\rm T}} \right)$,
with both $p$ and $\bar{\bar{\tau}}$ scaled with the characteristic value of the dynamic pressure $\rho'_0 U_0^2$. The power-law expressions $\mu=\rho D_{\sst{\rm T}}=T^{\sigma}$,
with $\sigma = 0.7$, are employed for the temperature dependence of the shear viscosity $\mu$ and thermal diffusivity $D_{\sst{\rm T}}$, both scaled with their feed-stream values. 

Equations~\eqref{cont}--\eqref{Z} 
must be integrated with appropriate conditions on the boundaries of the computational domain, which includes an outer cylindrical boundary with radius $r_\text{max} \gg 1$, with downstream and upstream boundaries located at $x=x_d$ and at $x=x_u$. The results corresponding to the most unstable mode were tested to be independent of the size of the computational domain, with the values $r_\text{max}=45$, $x_d=450$, and $x_u=-10$ selected for the computations shown below. The injector is assumed to be sufficiently long for the fuel flow to be fully developed, thereby giving $\vect{v}=2\,(1-r^2){\vect{e}}_x$ and $ {Z}=1$ in the fuel boundary upstream from the injector exit (\ie at $x=x_u$ for $0 \le r \le 1$). On the injector walls the solution satisfies the nonslip condition $\vect{v}=0$, together with the condition $\vect{n} \cdot \nabla  {Z} = 0$ corresponding to an impermeable wall, with $\vect{n}$ representing here the normal unit vector. To let the air enter or leave the computational domain as required to satisfy the development and the entrainment needs of the jet, a stress-free condition $-p \vect{n} +\bar{\bar{\tau}} \cdot \vect{n}/\Reyn=0$ is applied all around the outer air boundary. Air enters the flow field through the lateral boundary and through the upstream boundary, so that the condition $Z=0$ applies there, whereas $\vect{n} \cdot \nabla Z = 0$ must be used on the downstream boundary to allow for the evacuation of the combustion products. 

\section{The global linear stability analysis}

\label{The global linear stability analysis}

\subsection{The eigenvalue problem}

Introduction of the temporal normal-mode decomposition $(\vect{v},p,Z) = ({\bar{\vect{v}}},\bar{p},\bar{Z})+ \varepsilon ({\hat{\vect{v}}},\hat{p},\hat{Z}) e^{-\ui \omega t}$, involving the steady base flow $({\bar{\vect{v}}},\bar{p},\bar{Z})(\vect{x})$, the eigenfunctions $({\hat{\vect{v}}}, \hat{p}, \hat{Z})(\vect{x})$ multiplied by an arbitrarily small factor $\varepsilon$, and the complex angular frequency $\omega = \omr + \ui \omi$, leads to a set of nonlinear equations for the base flow (\ie the steady counterpart of \eqref{cont}--\eqref{Z}), to be integrated with the boundary conditions stated in the last paragraph of the preceding section. The associated linear equations for the perturbed flow
\begin{align}
	- \ui \omega \hat{\rho}  + \nabla \cdot \hat{\rho} \bar{\vect{v}} + \nabla \cdot \bar{\rho} \hat{\vect{v}} = & 0 , \\
	- \ui \omega \bar{\rho} \hat{\vect{v}} + \hat{\rho} \bar{\vect{v}} \cdot \nabla \bar{\vect{v}} + \bar{\rho} \hat{\vect{v}} \cdot \nabla \bar{\vect{v}} + \bar{\rho} \bar{\vect{v}} \cdot \nabla \hat{\vect{v}}  = & - \nabla \hat{p} - \frac{1}{\Frou} \hat{\rho} \vect{e}_x \notag \\
	+  \frac{1}{\Reyn} \nabla \cdot \Big[ \hat{\mu}  (\nabla \bar{\vect{v}}  &  + \nabla \bar{\vect{v}}^{\sst{\rm T}} ) + 
	   \bar{T}^\sigma \left(\nabla \hat{\vect{v}} + \nabla \hat{\vect{v}}^{\sst{\rm T}} \right) \Big],  \label{contlin} \\
		- \ui \omega \bar{\rho} \hat{Z} + \hat{\rho} \bar{\vect{v}} \cdot \nabla \bar{Z} +  \bar{\rho} \hat{\vect{v}} \cdot \nabla \bar{Z} +  \bar{\rho} \bar{\vect{v}} \cdot \nabla \hat{Z} = & \frac{1}{\Reyn \Pran} \nabla \cdot \left[ ( \hat{\mu} \nabla \bar{Z} + \bar{T}^\sigma \nabla \hat{Z} )  \right], \label{Zlin}
\end{align} 
arise from linearization of~\eqref{cont}--\eqref{Z}; these must be supplemented with $\hat{\rho}/\bar{\rho}=-\hat{T}/\bar{T}$ and $\hat{\mu}=\sigma \bar{T}^{\sigma-1} \hat{T}$, which follow from the equation of state and from the transport description, and with $\hat{T}=\gamma \hat{Z}/Z_S$ for $0\le \bar{Z} \le Z_S$ and $\hat{T}=-\gamma \hat{Z}/(1-Z_S)$ for $Z_S \le \bar{Z} \le 1$, which follow from~\eqref{rels}. Boundary conditions for~\eqref{contlin}--\eqref{Zlin} are $\hat{\vect{v}}=\hat{Z}=0$ in the fuel stream and  $\hat{\vect{v}}=\vect{n} \cdot \nabla \hat{Z}=0$ on the injector wall. On the air boundary, the stress-free condition for the perturbed flow reduces to $-\hat{p} \vect{n} +\left(\nabla \hat{\vect{v}} + \nabla \hat{\vect{v}}^{\sst{\rm T}} \right) \cdot \vect{n}/\Reyn=0$ on the upstream and lateral air boundaries, where $\hat{Z}=0$, and to $\sigma \gamma \, [(\bar{p}/\bar{T})(\hat{Z}/Z_S)- \hat{p}] \, \vect{n}+ \bar{T}^\sigma \left(\nabla \hat{\vect{v}} +  \nabla \hat{\vect{v}}^{\sst{\rm T}} \right) \cdot \vect{n}/\Reyn= 0$ on the downstream boundary, where $\vect{n} \cdot \nabla \hat{Z}=0$. Nontrivial solutions $({\hat{\vect{v}}}, \hat{p}, \hat{Z}) \ne 0$ are found for a discrete set of values of $\omega$, which is determined as an eigenvalue. The real part of $\omega$ is the frequency of the perturbation, defining a Strouhal number $\Stro = \omr/\upi$ (the ratio of the residence time $2a/U_0$ to the period of the oscillation); the imaginary part is the growth rate, which dictates whether the flame is globally stable ($\omi < 0$) or unstable ($\omi > 0$).

\subsection{Sample numerical results}

The base flow was integrated using a finite-element method with $P1$ elements for the pressure field and $P2$ elements for the remaining variables, combined with a Newton-Raphson root-finding algorithm; details of the discretization method, used for instance by \citet{Garnaud2013b}, can be found in~\cite{Hecht_FF}. The same finite-element formalism was employed to discretize the perturbed equations, resulting in a generalized eigenvalue problem that was solved using a shifted inverse power method \citep{ARP97}. 

The integrations explored in particular configurations with $2.08 \le S\le 9.66$ and moderately large values of $\Reyn$, for which the resulting flame height is much larger than the injector radius, as shown in figure~\ref{fig:sketchbaseflow} for the case $S=4.62$, $\Reyn = 100$, and $\Frou = 300$. A thick solid curve is used to denote the flame location, where $\bar{Z} = Z_S \simeq 0.178$ and $\bar{T}/(\gamma+1)= 1$. Besides isocontours of $\bar{Z}$, the plot includes streamlines, which serve to illustrate the motion of the air induced by the entrainment of the mixing layer surrounding the flame envelope.

Radial profiles of axial velocity $\bar{v}_x$ and normalized temperature $\bar{T}/(\gamma+1)$ are represented in figure~\ref{fig:sketchbaseflow} at different axial locations. Even for this relatively large Froude number, buoyancy is seen to accelerate the flow in the flame envelope, leading to the appearance of two inflection points in the velocity profile near the flame, additional to the inflection point associated with the shape of the initial velocity profile (the location of these inflection points is marked with a dot). As shown previously for mixing layers ~\citep{Soteriou1995} and low-density jets~\citep{Lesshafft2007}, the action of the baroclinic torque, induced in jet flames by the radial density gradient present in the near-flame region where the velocity profile displays inflection points, plays a key role in the development of a region of absolute stability \citep{Lingens1996b}, which in turn triggers the global oscillations. The rate at which the induced perturbations are convected away from this wave-maker region depends on the local value of the axial velocity, with smaller velocities favouring the development of absolute instabilities \citep{Lesshafft2010}.

\begin{figure}
    \centering
    \includegraphics[width=180pt]{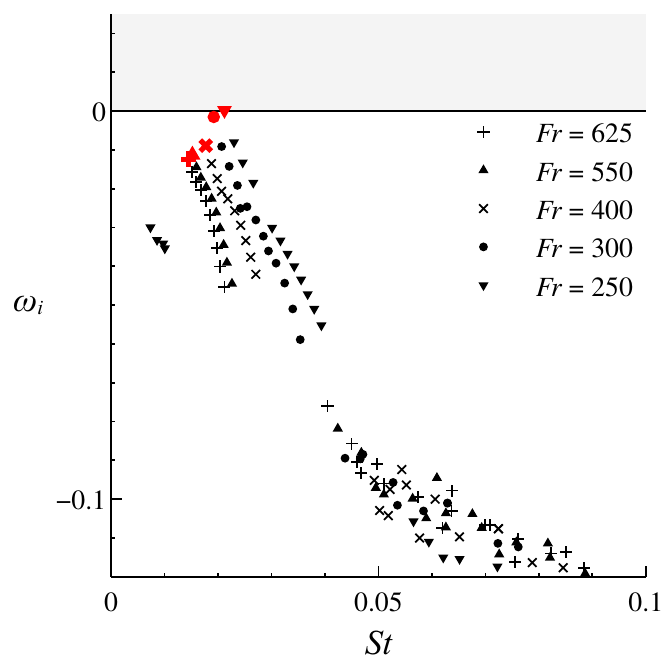}
    \llap{\parbox[b]{175pt}{(\textit{a})\\[160pt]}}
    \hfill
    \includegraphics[height=180pt]{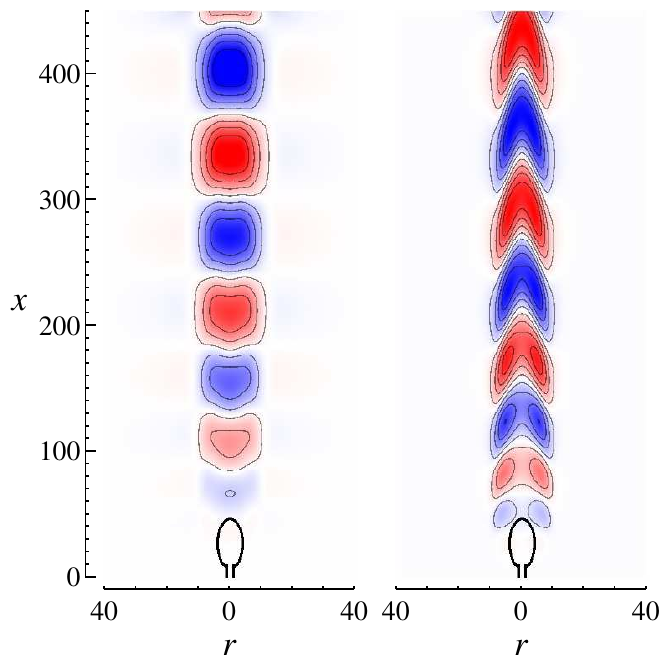}
    \llap{\parbox[b]{175pt}{(\textit{b})\\[160pt]}}
    \caption{(a) Eigenvalue spectra for $\Pran=0.7$, $S=4.62$, $\gamma=6$,
        $\Reyn = 100$, and $\Frou = (250,300,400,550,625)$. 
        (b) Real part of the streamwise velocity $\hat{v}_x$
        (left) and mixture fraction $\hat{Z}$ (right) for the
        eigenfunctions of the most unstable mode with $\Reyn = 100$ and
        $\Frou = 300$.}
    \label{fig:spectrumandeigf}
\end{figure}

Figure~\ref{fig:spectrumandeigf}(a) shows the eigenvalue spectra computed for $\Reyn=100$ and different values of $\Frou$. For all cases, the most unstable
eigenmode is indicated with a bigger symbol in red. Decreasing the Froude number is seen to destabilize the flow, so that for $\Frou=300$ the growth rate $\omi$ of the most unstable mode is still negative, but it is already positive for $\Frou=250$. 
For completeness, eigenmodes corresponding to the subcritical case $\Frou=300$ are plotted in figure~\ref{fig:spectrumandeigf}(b). As can be seen, both the radial extent of the eigenmodes and their wavelength scale with the flame dimensions. Although the length $x_d=450$ of the computational domain was not long enough to capture the downstream decay of the eigenmodes, the associated values of $\omega$ for the most unstable mode were seen to be independent of $x_d$ provided that $x_d>400$, as verified in a series of computations.

\subsection{Transition diagrams}

\begin{figure}
\centering
    \includegraphics[width=320pt]{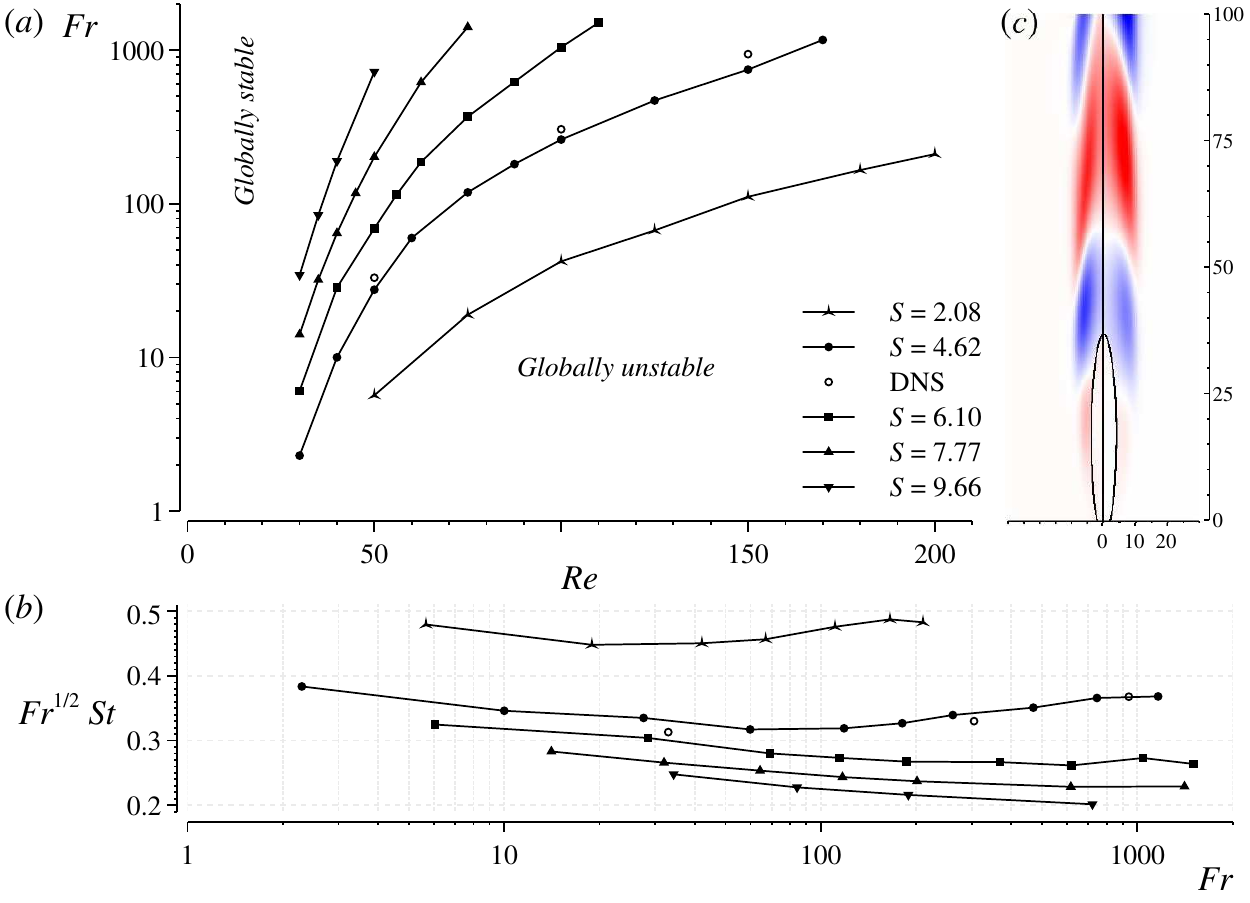}
    \caption{(a) Transition diagram in the $\Reyn-\Frou$ parametric plane for different values of $S$, with the (b) accompanying panel showing the variation with $\Frou$ of the nondimensional frequency $\Frou^{1/2} \Stro = (\omr/\upi)/\sqrt{g/a}$ at the margin of instability and the empty symbols in both plots representing DNS predictions for $\Reyn = (50, 100, 150)$. As explained in the text, (c) a comparison for $S=4.62$, $\Reyn=100$, and $\Frou=261.40$ of the eigenmode $\hat{T}(\vect{x})$ (right-hand side) with a snapshot extracted from the DNS results (left-hand side).}
    \label{fig:transdiag}
\end{figure}

Marginal conditions were determined by linear interpolation of the results of stability spectra computed for given values of $S$ and $\Reyn$ and decreasing values of $\Frou$ (including stable and unstable cases), giving the transition diagrams and accompanying frequencies shown in figure~\ref{fig:transdiag}. The resulting marginal curves serve to assess effects of fuel-feed dilution and of molecular transport. Increasing $\Reyn$ for a given value of $S$ is seen to have a destabilizing effect, in that the global instability sets in at a higher value of $\Frou$, in agreement with recent observations for low-density jets \citep{Coenen2016}. Conversely, fuel-feed dilution (\ie decreasing values of $S$) tends to stabilize the flow, a result that can be explained by noticing that dilute flames sit closer to the axis, where the downstream convective rate of the perturbations is higher, thereby hindering the development of a region of absolute instability and resulting in smaller critical values of $\Frou$.

The large variations in critical values of  $\Frou$ observed in figure~\ref{fig:transdiag} would be considerably reduced should the characteristic scales of the flame, rather than those associated with fuel injection, be used in defining the relevant Froude number \citep{LVL_2015}. Thus, with $\Reyn \gg 1$ and $S\gg 1$ the flame length is of order $S \Reyn a$. At these distances, the jet velocity has decreased to values of order $U_0/S$, which must be compared with the buoyancy-induced velocity $g S \Reyn a$, the square of their ratio giving $\Frou/(S^3 \Reyn)$ as the relevant Froude number for jet flames. Inspection of the results in figure~\ref{fig:transdiag} reveals that this alternative definition would result in a transition diagram with less pronounced variations of the critical Froude number over the range of conditions explored.

For the range of Reynolds numbers explored in figure~\ref{fig:transdiag}, the effective Froude number accounting for the residence time in the flame region $\Frou/(S^3 \Reyn)$ is somewhat smaller than unity, corresponding to jet flames with significant buoyancy effects. Consequently, the dynamics of the resulting oscillations at the margin of stability is characterized by the buoyancy time $\sqrt{a/g}$, rather than by the residence time $a/U_0$ employed initially in nondimensionalizing the problem. This scaling is tested in figure~\ref{fig:transdiag}, where the frequency is represented in terms of $(\omr/\upi)/\sqrt{g/a}=\Frou^{1/2} \Stro$. As can be seen, for each value of $S$ the resulting frequencies change only by about 10\% over the whole range of Froude numbers explored in the figure, thereby demonstrating the prevalence of the buoyancy scaling. This is in agreement with the experimental observations of \cite{Durox1997} and \cite{Sato2000}, who found that $\Stro \sim \Frou^{-1/2}$.

The buoyancy-dominated flickering mode observed here is markedly different from that corresponding to buoyancy-free light jets and flames, for which the resulting frequencies scale with $a/U_0$, with associated eigenmodes scaling with the jet radius, rather than with the flame length. This alternative buoyancy-free mode, which must become dominant at sufficiently high Froude numbers (and sufficiently high accompanying Reynolds numbers), was not observed in the computations carried out here. For the Poiseulle exit-velocity profile considered in our work, preliminary computations for $\Frou = \infty$ and $S=6.10$ indicated that nonbuoyant flames are globally stable up to the largest Reynolds number considered ($\Reyn = 1000$). A critical Reynolds number exceeding 1000 for buoyancy-free jet flames is consistent with previous results concerning the influence of the exit velocity profile on the stability of light jets \citep{Hallberg2006}. Smaller critical values of the Reynolds number are expected, for instance, for nearly uniform profiles, encountered with shorter fuel injectors.

\section{Comparison with DNS results}
\label{sec:comp}

The predictions of the stability analysis at the margin of stability were compared with DNS results obtained with a time-dependent axisymmetric code \citep{JC_2016} using the same grid employed in the global stability computations. The numerical simulations at three different points along the marginal curve for $S=4.62$, namely, $(\Reyn,\Frou)=(50,26.60)$, $(\Reyn,\Frou)=(100,261.40)$, and $(\Reyn,\Frou)=(150,745.64)$, yielded periodic solutions with small amplitude. The associated Strouhal numbers, $\Stro =(0.0628,0.0200,0.0132)$, obtained by fitting the oscillations of the numerical solutions to a sinusoidal function, were seen to be in excellent agreement with the values $\Stro =(0.0638,0.0190,0.0134)$ predicted by the stability analysis. The agreement extends to the morphology of the flickering mode, as can be seen in the inset of figure~\ref{fig:transdiag}, which compares the eigenmode $\hat{T}(\vect{x})$ corresponding to $(\Reyn,\Frou)=(100,261.40)$ with the near-critical DNS results, the latter obtained by subtracting the time-averaged temperature from the instantaneous distribution $T(\vect{x};t^*)$, with the time $t^*$ appropriately selected to minimize the observed differences. As can be seen, there exist excellent agreement not only in the predicted wavelength but also in the shape of the cells representing the traveling rollers.

As mentioned above, for the parametric values corresponding to the marginal conditions of the stability analysis, the DNS results were seen to exhibit small oscillations of nonnegligible amplitude. Additional computations for increasing values of $\Frou$, resulting in periodic solutions with decreasing amplitude, were performed to determine the marginal curve predicted by the DNS results. The transition to the flickering state is governed by a supercritical Hopf bifurcation. Correspondingly, with the Froude number being the relevant bifurcation parameter for fixed values of $S$ and $\Reyn$, the amplitude of the oscillations near the margin of stability is expected to exhibit the proportionality $A^2 \sim (\Frou - \Frou^*)$ \citep[][\S~27]{Landau1959}, where $\Frou^*$ is the critical value of $\Frou$. This is illustrated in figure~\ref{fig:transdiag}(a), which shows the squared amplitude of the mixture-fraction oscillations along the axis at four different downstream locations, as obtained in numerical simulations for $\Reyn = 50$, $S = 4.62$, and decreasing values of $\Frou$. Extrapolating the corresponding results to zero amplitude provides the critical value $\Frou^*$ of $\Frou$, giving for instance  $\Frou^*=(33,305,940)$ for $\Reyn=(50,100,150)$. These values are compared in figure~\ref{fig:transdiag} with the values $\Frou^*=(26.60,261.40,745.64)$ corresponding to the linear stability analysis.

The direct numerical simulations also indicate that near the margin of stability there exists a linear dependence of the oscillation frequency on the Froude number. As shown in figure~\ref{fig:transdiag}(b), the observed frequency is identical at all four locations---confirming the global nature of the flicker instability---with the critical value $\Stro^* = 0.054$ approached as $\Frou \rightarrow \Frou^* \simeq 33$. This is to be compared with the value $\Stro = 0.0628$ predicted by the global stability analysis at the corresponding critical Froude number $\Frou = 26.60$. The differences observed, for both critical Froude numbers and associated frequencies, whose relative magnitude is on the order of 20\% in the range of Reynolds numbers investigated, may be attributed to the fact that disturbances experience very large gains in slightly subcritical settings, leading to a substantial amplification of small numerical noise in the DNS integrations that results in the larger critical values of $\Frou$ shown in figure~\ref{fig:transdiag}. Clearly, the origin of the observed discrepancies warrants further investigation in future work.

\begin{figure}
\centering
    \includegraphics[width=240pt]{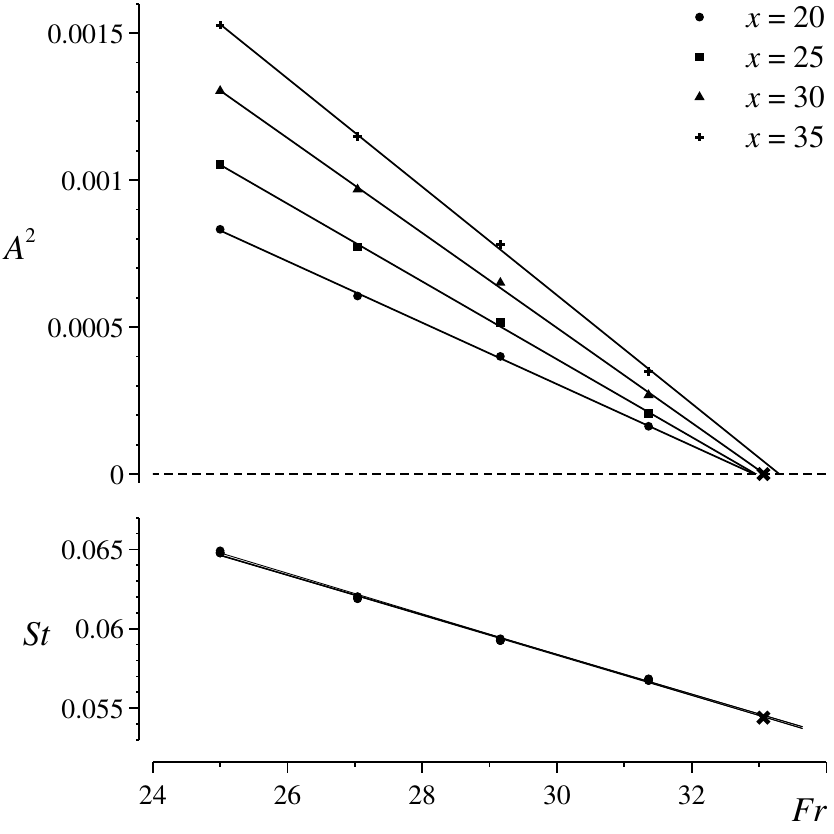}
    \caption{DNS computations of the (a) squared amplitude of oscillations, $A^2$, and (b) Strouhal numbers $\Stro$, measured at different downstream axial locations, $x = 20, 25, 30$ and $35$ for $\Reyn = 50$ and $S = 4.62$. The solid lines in the upper part of the plot represent linear fits of the points $(\Frou, A^2)$ in order to determine the critical Froude number, $\Frou^*$, for which the amplitude of the oscillations is zero. Upon extrapolation, a critical value $\Frou^* = 33.065$ was found, with an associated Strouhal number $\Stro^* = 0.0544$, marked with a cross, whereas the Strouhal number for the conditions predicted by the global stability analysis was found to be $\Stro = 0.0628$.}
    \label{fig:datadns}
\end{figure}

\section{Comparison with a local stability analysis}
\label{sec:local}

\begin{figure}
\centering
    \includegraphics[width=384pt]{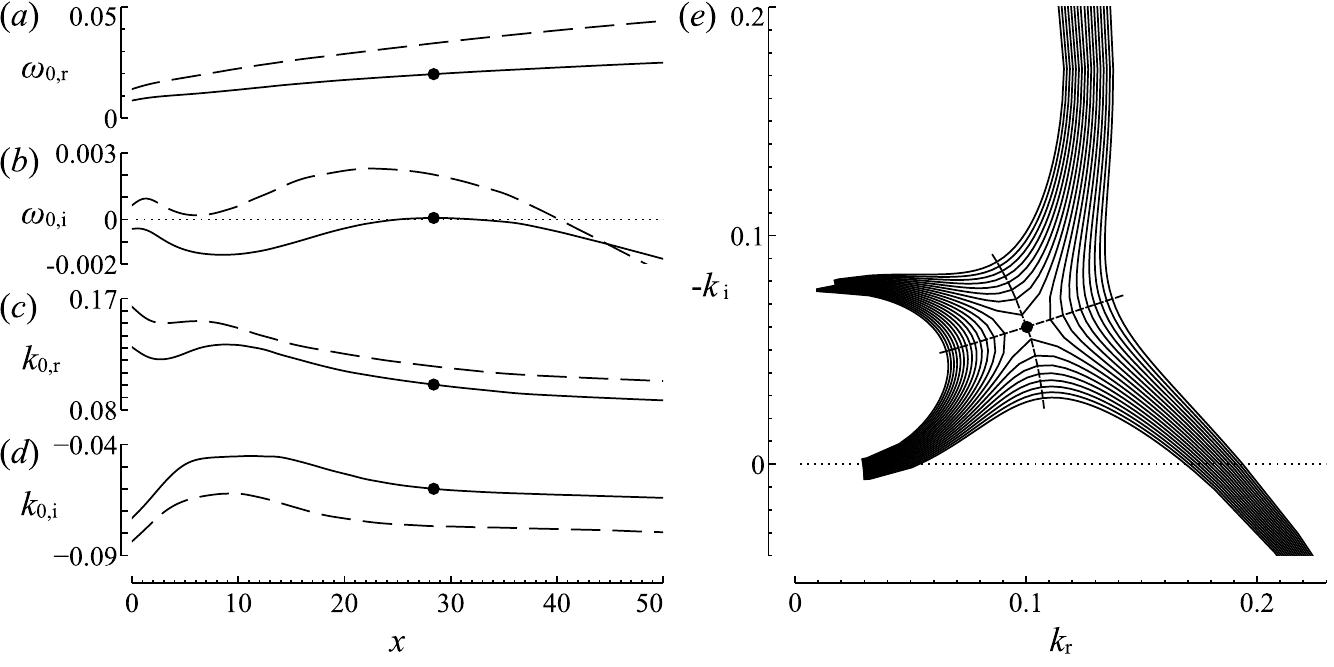}
    \caption{
    (a)--(d) Downstream evolution of the local spatio-temporal stability properties for $S=6.1$ and $\Reyn=75$, with $\Frou=800$ (solid line) and $\Frou=375$ (dashed line). (e) Location of the saddle point $(\omo,\ko)$ in the complex $k$-plane at the downstream position $x=28.4$, as indicated in figures (a)--(d) by the dots; the solid lines are spatial branches with a constant value of $\omi$; the dashed lines are lines of constant $\omr=\omor$.%
    }
    \label{fig:local}
\end{figure}

As explained in \citet{Huerre1990}, for slender flows there exists a close relationship between the evolution of the local stability characteristics at each streamwise position $x$, and the global instability properties of the flow. However, this relationship depends crucially on the requirement that the wavelength $\lambda$ be much smaller than the typical evolution length scale $L$ of the basic flow; and, quoting \citet{Huerre1990}, ``A breakdown of this assumption would preclude any possible connection between local and global instability properties''. For the diffusion flame presented in figure~\ref{fig:transdiag}, it can be seen that the wavelength~$\lambda$ of the global instability is comparable to the flame height, which characterizes the spatial evolution of the base flow. Therefore, the conditions needed for applicability of the local spatio-temporal analysis are not satisfied, which may result in significant inaccuracies in inferred predictions of global instability properties. This aspect of the problem is to be investigated here. Specifically, we shall study the downstream evolution of the local spatio-temporal stability properties of the base flow used earlier for the global stability analysis. We begin by formulating the local stability analysis, and then show results for the case $S = 6.1$ and $\Reyn = 75$, with $\Frou=375$ and $\Frou=800$. In analyzing the results it is worth bearing in mind that the global instability analysis predicts a critical Froude number $\Frou = 368$ for $S = 6.1$ and $\Reyn = 75$, so that the flow should be globally stable under these conditions.

At each downstream position $x$, the basic flow is assumed to be locally parallel, with radial profiles of velocity $\vvb(r)=(\vxb(r), 0)$ and mixture fraction $\Zb(r)$; small perturbations are introduced as normal modes $[\vxt(r), \ui \vrt(r), \pt(r), \Zt(r)] \exp[\ui(k x - \omega t)]$, with complex axial wavenumber $k = \kr + \ui \ki$ and complex angular frequency $\om = \omr + \ui \omi$. Here $k$, $\om$, and $t$ are nondimensionalized using $a$ and $U_0$. In appendix~\ref{sec:applocalform} it is shown how substitution of the normal modes into the equations of motion~\eqref{cont}--\eqref{Z}, linearized around the steady base flow, yields the system of ordinary differential equations \eqref{eq:localcont}--\eqref{eq:localZ} that, together with the boundary conditions \eqref{eq:localbcinf}--\eqref{eq:localbcaxis}, provides a generalized eigenvalue problem. The local stability properties can be obtained by solving the latter, whereby eigenfunctions $\vxt(r)$, $\vrt(r)$, $\pt(r)$, $\Zt(r)$ only exist if $k$ and $\omega$ satisfy a dispersion relation $D(k,\om; \Reyn, \Frou, S, \gamma, \ldots, \vxb, \vrb, \pb, \Zb)=0$. In the present section we are concerned with the absolute or convective character of the instability. Therefore we need to find the spatio-temporal instability modes with zero group velocity, \ie modes for which $\ud \om / \ud k = 0$. The growth rate $\omoi$ of these is called the absolute growth rate and determines whether the instability is convective, $\omoi<0$, or absolute, $\omoi>0$. The condition $\ud \om / \ud k = 0$ is equivalent to the existence of a double root, or saddle point, in the complex $k$-plane, $\left. \pfi{D}{k} \right|_{k = k_0} = 0$. Among all the saddle points that may exist, only the one with the largest value of $\omoi$, while satisfying the Briggs--Bers criterion, determines the large-time impulse response of the flow \citep[see, for instance,][and references therein]{Huerre2000}. The numerical method used to determine $(\omo,\ko)$ is described in appendix~\ref{sec:applocalnum}.%

Figures~\ref{fig:local}$(a)$--$(d)$ show the downstream evolution of the spatio-temporal stability properties for the case $S=6.1$ and $\Reyn=75$, with two values of the Froude number: $\Frou=800$ (solid lines), and $\Frou=375$ (dashed lines). The location of the saddle point $k = \ko$ in the complex $k$-plane is shown in figure~\ref{fig:local}$(e)$, where the solid lines indicate spatial branches of constant $\omi$, and the dashed lines have a constant value of $\omi=\omoi$. It can be seen how for $\Frou \lesssim 800$ a pocket of absolute instability emerges around $x = 28$, with absolute frequency $\omor = 0.02$ ($\Stro=0.006$) and wavelength $\lambda_0 = 2\upi/\kor = 63$. In numerical simulations of weakly nonparallel heated jets, the appearance of such a pocket of absolute instability was shown to destabilize the nonlinear global mode responsible for the self-excited behaviour \citep{Lesshafft2007a}. Moreover, at criticality, the corresponding global frequency was found to coincide with the value given by the local stability analysis at the downstream position where the character of the instability changes from convective to absolute, in agreement with the theory developed by \cite{Pier1998} for weakly nonparallel flows.

The spatio-temporal stability analysis therefore predicts the flow to be globally unstable for $\Frou \lesssim 800$, with a frequency at the margin of instability such that $\Stro=0.006$. These predictions differ significantly from those of the global stability analysis, which gives a critical Froude number $\Frou = 368$ with an associated Strouhal number $\Stro \simeq 0.014$. These departures can be attributed to the failure of the condition $\lambda \ll L$ needed for applicability of the quasi-parallel analysis. Similar overpredictions in the growth rate of the perturbations have been reported in previous comparative studies of local/global stability analyses for wakes \citep{Juniper2011}.  

A pocket of absolutely unstable flow, away from boundaries, was also found by \cite{Qadri2015} in the context of nonbuoyant flames for their ``mode B''. As in the present work, they found this region of local absolute instability to lie at the basis of the excitation of a global low-frequency flickering mode. In the buoyancy-free configuration analyzed by \cite{Qadri2015} the density of the fuel jet upstream from the lifted flame is significantly lower than that of the surrounding atmosphere, causing a second instability mode (``mode A'') to be present in their analysis, with a region of absolute instability that starts at the outlet of the jet, similar to that found by \citet{Coenen2012} in the context of light jets.%

\section{Conclusions}
\label{sec:conclusions}

The present investigation has employed, for the first time, a global stability analysis to study the buoyancy-induced flickering of jet diffusion flames as a hydrodynamic global mode. The paper provides the parametric dependence of the critical conditions at the onset of instability as well as the morphology and frequency of the resulting oscillatory modes, giving predictions in fair agreement with results of direct numerical integrations. 

While the simplified model employed here contains the fundamental underlying physics involved in the flickering phenomenon, additional effects should be investigated in future work. For instance, influences of shapes of jet-velocity profiles, including interactions of the different instabilities observed previously \citep{Roquemore1988}, could be investigated by incorporating the boundary-layer thickness as an additional parameter, as done in previous spatio-temporal stability analyses of light jets \citep{Coenen2008}. Preferential diffusion effects, associated with light and heavy fuel molecules, could be addressed in the infinitely fast reaction limit by using  coupling-function formulations accounting for reactant Lewis numbers that differ from unity~\citep{Li_91}. Consideration of finite-rate effects would be needed to examine the stability characteristics of lifted flames, studied in previous work \citep{Qadri2015} under buoyancy-free conditions. While the present work pertains to laminar flames, the global instability analysis could also be applied to turbulent conditions, with the steady base flow obtained for instance by time averaging results of large-eddy simulations, as done earlier in connection with local spatiotemporal analyses of jet flames \citep{SI_2014}.

While the mode identified here is buoyancy-dominated, resulting in frequencies that scale with $(g/a)^{1/2}$, the dynamics at sufficiently large Froude numbers is expected to be controlled by a different mode, with frequencies that scale with $U_0/a$, similar to those observed in light jets \citep{Hallberg2006}. Our preliminary computations indicate that the investigation of the transition between the buoyancy-dominated and the momentum-dominated instabilities will require consideration of much higher Froude numbers, with associated critical Reynolds numbers exceeding $\Reyn=1000$. The associated global instability computation is expected to experience difficulties associated with the existence of resonance modes caused by spurious feedback from the outflow boundary, encountered earlier in the analysis of jets \citep{Garnaud2013a}. 

Our analysis indicates that the streamwise wavelength of the dominant instability mode scales with the flame height. Correspondingly, the assumption of quasi-parallel flow, necessary to ensure the predictive capability of local stability analyses, does not hold in buoyant jet diffusion flames, resulting in associated predictions of critical Froude numbers at the margin of instability that are off by a factor exceeding two. This finding further underscores the utility of global instability analysis for investigation of buoyancy-induced flickering instabilities.

The global linearized approach opens up a range of possibilities for further studies. For instance, the computation of the adjoint modes---with the discretized Navier--Stokes operators at one's disposal, the discrete adjoint can be obtained in a straightforward manner by solving the conjugate-transposed eigenvalue problem---readily permits a structural sensitivity analysis in the sense of \cite{Giannetti2007}. Hereby the sensitivity of the eigenvalue with respect to `internal feedback' mechanisms is obtained by measuring the local overlap between the direct and the adjoint eigenfunctions. It is argued that flow regions where this measure is large contribute strongly to the eigenvalue selection and thus represent the `wavemaker' of the eigenmode.  Another interesting concept is the sensitivity to a steady body force or to modifications in the base flow \citep{Marquet2008}. This is particularly relevant in the context of passive control techniques, such as the introduction of an adequately positioned control cylinder to stabilize the flame flicker \citep[see, for instance][]{Toong1965}. These sensitivity analyses have been recently applied to nonbuoyant lifted flames \citep{Qadri2015}.

Finally, linear nonmodal stability techniques may be applied to investigate the discrepancy between the onset of instability predicted by the global stability analysis and that obtained by DNS. A similar difference has recently been encountered by \cite{Coenen2016} in low-density jets when comparing the results of a global stability analysis with the experimental observations of \cite{Hallberg2006}. In that regard, the computation of the pseudospectrum \citep{Trefethen2005} of the linearized Navier--Stokes operator can show if non-normality plays a role. For the low-density jet, a very large gain in the frequency response \cite[see also][]{Garnaud2013b} was found, even for Reynolds number substantially smaller than the critical value. These aspects should be investigated for buoyant jet diffusion flames in future work.%

\section*{Acknowledgements}
Norbert Peters pointed out the need for the present analysis in his seminal paper with John Buckmaster published thirty years ago~\citep{Buckmaster1986}. It is with great sorrow that we received the news of his passing last year. This paper is devoted to his memory in gratitude for his many outstanding contributions to Combustion Science.

The constructive comments of one anonymous referee have led to substantial improvements of the paper and are gratefully acknowledged. This work was supported by the Spanish MCINN through project \# CSD2010-00010 and by the Spanish MINECO through project \# DPI2014-59292-C3-1-P.


\oneappendix

\section{Details of the local stability analysis}

\subsection{Stability equations}
\label{sec:applocalform}

To obtain the stability equations, the normal modes $\{\vxt(r), \ui \vrt(r), \pt(r) \Zt(r)\} \exp[\ui(k x - \omega t)]$ are substituted into the equations of motion \eqref{cont}--\eqref{Z}, linearized around the base flow $\{\vxb(r), \vrb(r), \pb(r), \Zb(r) \}$, yielding the system of ordinary differential equations
\begin{align}
    & - \om \rhot + k \rhob \vxt + \rhob (\vrt' + \vrt/r) + \rhob' \vrt + k \vxb \rhot = 0,
    \label{eq:localcont} \\
    & \ui \rhob \left( - \om \vxt + k \vxb \vxt + \vxb' \vrt \right)
        = - \ui k \pt - \Frou^{-1} \rhot \notag \\
    &\qquad + \Reyn^{-1} \left[ \mub (\vxt'' + \vxt'/r - k^2 \vxt)
                        + \mub' (\vxt' - k \vrt)
                        + (\vxb'' + \vxb'/r ) \mut + \vxb' \mut' \right],
    \label{eq:localmomx} \\
    & \ui \rhob \left( - \om \vrt + k \vxb \vrt \right)
        = \ui \pt' \notag \\
    &\qquad + \Reyn^{-1} \left[ \mub (\vrt'' + \vrt'/r - \vrt/r^2 - k^2 \vrt)
                        + \mub' (\vrt' - \vrt/r - k \vxt)
                        + k \vxb' \mut \right],
    \label{eq:localmomr} \\
    & \ui \rhob \left( - \om \Zt + k \vxb \Zt + \Zb' \vrt \right)
        = \notag \\
    &\qquad + {(\Reyn\,\Pran)}^{-1} \left[ \mub (\Zt'' + \Zt'/r - k^2 \Zt)
                                         + \mub' \Zt'
                                         + (\Zb'' + \Zb'/r ) \mut + \Zb' \mut' \right],
    \label{eq:localZ}
\end{align}
where the prime indicates differentiation with respect to~$r$. Note that
\begin{align}
    \Tb & =
    \begin{cases}
        1 + \gamma \Zb/\Zs         & \text{for $0 \le \Zb \le \Zs$}, \\
        1 + \gamma (1-\Zb)/(1-\Zs) & \text{for $\Zs \le \Zb \le 1$},
    \end{cases} \\
    \rhob & = \Tb^{-1}, \\
    \mub  & = \Tb^\sigma,
\end{align}
and
\begin{align}
    \Tt & =
    \begin{cases}
         \gamma \Zt/\Zs     & \text{for $0   \le \Zb \le \Zs$}, \\
        -\gamma \Zt/(1-\Zs) & \text{for $\Zs \le \Zb \le   1$},
    \end{cases} \\
    \rhot & = -\Tb^{-2} \Tt, \\
    \mut  & = \sigma \Tb^{\sigma-1} \Tt.
\end{align}

The stability equations are accompanied by suitable boundary conditions. In the far field, all perturbations must vanish,
\begin{equation}
    (\Zt, \vxt, \vrt, \pt) \to 0 \ \text{as $r \to \infty$},
    \label{eq:localbcinf}
\end{equation}
whereas at the centerline, a vanishing azimuthal dependence of the perturbations as $r \to 0$ may be imposed \citep{Batchelor1962}, leading to
\begin{equation}
    \vrt = \vxt' = 0 \ \text{and} \ (\Zt, \vxt, \pt) \ \text{finite} \ \text{at $r = 0$}.
    \label{eq:localbcaxis}
\end{equation}
Note that a Taylor expansion of~\eqref{eq:localmomx}--\eqref{eq:localZ} around the centerline yields
\begin{align}
    \mub \vxt' + \mut \vxb' & = 0,
    \label{localbcaxistaylorx} \\
    \Reyn^{-1} k ( \mub \vxt' + \mub' \vxt ) - 3 \Reyn^{-1} \mub \vrt''/2- \ui \pt' & = 0,
    \label{localbcaxistaylorr} \\
    \Zb' \mut + \mub \Zt' & = 0.
    \label{localbcaxistaylorZ}
\end{align}

\subsection{Numerical method}
\label{sec:applocalnum}

The numerical method that is employed to obtain the saddle point $(\omo,\ko)$ is identical to that of \citet{Coenen2012}. A quadratic Taylor expansion of $\om(k)$ around $(\omo,\ko)$ permits the employment of a Newton-Raphson root-finding algorithm, whereby at each iteration the temporal eigenvalue problem must be solved \citep{Deissler1987}. To that end, a spectral collocation method is used with a function $\xi = [r_\text{c} - r(1+2r_\text{c}/r_\text{max})]/(r_\text{c}+r)$ that maps the $N$ collocation points from the Chebyshev interval $-1 \leq \xi \leq 1$ to the physical domain $0 \leq r \leq r_\text{max}$ \citep{Khorrami1989}. Values $r_\text{c} = 3$, $r_\text{max}=50000$ and $N = 300$ are found to be adequate. For more details on the numerical method, we refer to \citet[][Appendix B]{Coenen2012}.


\end{document}